\documentclass[aps,prl,twocolumn,superscriptaddress]{revtex4-1}

\usepackage{graphicx}
\usepackage{dcolumn}
\usepackage{amsmath,amsthm}
\usepackage{amsfonts}
\usepackage{amssymb}
\usepackage{amssymb}
\usepackage{epstopdf}
\usepackage{times}
\usepackage{mathrsfs}
\usepackage{color}
\usepackage{gensymb}
\usepackage{times}
\usepackage{subfigure}
\usepackage[]{graphicx}
\usepackage{amssymb}
\usepackage{amsmath}
\usepackage{bm}
\DeclareGraphicsExtensions{.PNG,.jpg,.pdf,.gif,.eps}
\usepackage{amsmath,amssymb,amsthm,amsfonts,eucal}
\usepackage[english]{babel}

\usepackage[utf8]{inputenc}
\usepackage[T1]{fontenc}

\newcommand{\de}{\partial}

\newcommand{\eps}{\varepsilon}

\DeclareMathOperator{\sech}{sech}

\begin{document}

\title{Landau theory of  bending-to-stretching  transition}
\author{O.	U. Salman}
\affiliation{CNRS, LSPM UPR3407, Université Paris 13, Sorbonne Paris Cité, 93430 Villetaneuse, France}
\author{G.Vitale}
\affiliation{Laboratoire de Mécanique des Solides, CNRS-UMR 7649, Ecole Polytechnique, Route de Saclay, F-91128 Palaiseau Cedex, France}
\author{L. Truskinovsky}
\affiliation{PMMH, CNRS - UMR 7636 PSL-ESPCI, 10 Rue Vauquelin, 75005 Paris, France}
\date{\today}
\begin{abstract}
Transition from  bending-dominated to  stretching-dominated  elastic response  in semi-flexible  fibrous networks  plays  an important role in the mechanical behavior  of cells and tissues.   It is   induced by  changes in network connectivity and  relies on   construction of new cross-links. We propose a simple continuum  model of this transition with  macroscopic strain playing the role of order parameter. An unusual feature of this Landau-type theory is that it is based on a single-well potential. We predict  that  bending-to-stretching transition   proceeds  through  propagation of  the localized fronts separating  domains with affine and non-affine elastic response.  
\end{abstract}
 
\maketitle
 
Typical   force transmitting systems in cellular biology can be viewed at the microscale as  networks of cross-linked semi-flexible fibers which  respond to mechanical loading by both \emph{stretching} and \emph{bending} \cite{PhysRevLett.91.108103,PhysRevE.68.061907,onck2005,Kang2009-fb,Broedersz2014-tv,Pegoraro2017-ur}. One of the most striking features of such  'materials' is the loading-induced  transition from   non-affine, bending-dominated elasticity, to   (almost) affine, stretching-dominated  elasticity \cite{buxton2007,PhysRevE.87.042601,PhysRevLett.108.078102}. This transition is accompanied by the anomalous growth  of elastic moduli  and is usually  linked to  the increase of  the cross-linker density \cite{Feng2016-vq}. 
 
In highly connected dense networks the stretching stiffness  dominates because they cannot be deformed without either elongation or shortening of the links;  in less dense, under-constrained networks,  classical rigidity  is lost due to the appearance of floppy modes and softer,  bending elasticity  becomes responsible for the overall stiffness \cite{Head2003-hp,Das2007-et,Broedersz2011-vf}.  The bending-to-stretching (BS) transition   was successfully simulated in 2D and 3D athermal microscopic  models, and  it was found that 
  a continuous  crossover between the two  regimes takes the form of   a   highly heterogeneous   \emph{coexistence} between bending (B) and stretching (S) dominated \emph{phases} \cite{PhysRevLett.91.108103,PhysRevE.68.061907,onck2005,buxton2007,PhysRevE.87.042601,PhysRevLett.108.078102}. 
  
Despite these successes in microscale modeling, the  fundamental understanding of the  BS  transition at the macroscopic  level  is still lacking. The development of a coarse-grained model of this phenomenon  will facilitate the  continuum  modeling   of   cellular scale phenomena ~\cite{Meng2018-hz,Prost2015-tt,Bernheim-Groswasser2018-xr,Burkel2017-bx,Taloni2015-an}   and advance the design of the artificial  meta-materials with under-connected  network architecture \cite{Ashby2006-pr,Broedersz2014-tv,Rocklin2017-pz}.

In this Letter we develop a prototypical Landau-type  theory of   the strain induced BS transition. We build on the idea  that   cross-linked networks have  the ability  to internally rearrange in response to the applied deformation through fiber rotation   \cite{Onck2005-me,Heussinger2006-pe,Wen2013-gc}  and that new cross-links   can  form  in this process  \cite{Gardel2004-qc,Pegoraro2017-ur}.    Our main result is the \emph{regime diagram}  showing how the dominating   deformation mode is controlled by  the applied strain and the dimensionless ratio of the internal and external length scales.

\begin{figure}[t]
\includegraphics[scale=0.22]{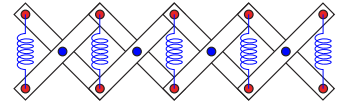}
\caption{Floppy network  reinforced by  elastic bonds  (represented by vertical springs) which can   disengage   at sufficient longitudinal shortening of the structure. Instead, horizontal stretching of the structure can lead to the engagement of the bonds.}
\label{fig1}
\end{figure}
Our approach is deliberately minimalistic. As a prototype  of a  semi-flexible network,   we use a  pantographic structure with freely rotating cross-links, as in a collapsible arm of wall mounted mirror \cite{Giessen:2011aa}. The crucial assumption is that this floppy mechanical system can be  stabilized  by    elastic bonds  whose role is   to  ensure   rigidity when they are intact, see  Fig.~\ref{fig1}.  

Suppose that the initial state is chosen in such a way that  all vertical  springs in Fig.~\ref{fig1} are disengaged and  the system is  under-constrained \cite{Maxwell1864-or,Calladine1978-zc}.   If  such structure is   stretched,   the geometrical constraints  force the system to contract in the vertical direction  which  can lead to the rebuilding of the bonds. As a result,  an  under-constrained system  transforms into an  over-constrained one.

We assume that the floppy structure  itself is built of  inextensible but flexible beams connected through pivots.  It is known that the macroscopic elastic response of the  unreinforced pantographic  structure shown in Fig.~\ref{fig1} is  B-dominated \cite{alibert:hal-00993920};  more complex   examples   can be found in the theory of  high contrast  elastic composites  \cite{cherednichenko2006,boutin2013,camar2003}. In  the continuum representation of such   systems the  non-local (higher order) elasticity appears already at the leading order in the homogenization limit which  leads to elasticity theories dominated by an internal length scale (as in liquid crystals  \cite{Chaikin1995-xw}).  
 
\begin{figure}[t]
\includegraphics[scale=0.21]{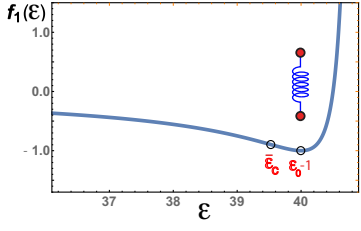}
\caption{The flipped Lennard-Jones type potential $f_1$ with $\eps_0=41$.  At the spinodal limit $\de^2 f_1( \bar\eps_c) = 0$.}
\label{fig2}
\end{figure}

We can then model the discrete structure shown in  Fig. \ref{fig1} as a continuum bar whose classical elastic energy 'softens' in compression due to   breaking of the reinforcing springs. An  additive quadratic strain gradient  term in the energy density   can be used as a proxy for  the higher order (non-classical)  elasticity of the pantographic frame. With  macroscopic strain    playing the role of the      order parameter \cite{golubovic1989},  the ensuing model takes the form of  a   Ginzburg-Landau  (GL)  theory  \cite{Fracturett}.  It is characterized, however, by  an unusual   vertically flipped  Lennard-Jones (Morse)  type potential, see Fig. \ref{fig2}.

We use this continuum model to show that  the quasi-statically driven BS transition proceeds  through nucleation and propagation of  the  fronts separating  domains with affine and non-affine elastic response. At such fronts,  the  connectivity of the network  changes  and they can be interpreted as the (degenerate)  domain walls. 
In contrast to the mixed  BS   states,  the pure B and S states are homogeneous.
We show, however, that  in the B states the affinity,   imposed by the  weak gradient elasticity, is not robust.
   
To mimic realistic systems we also consider the GL model with a  constraining elastic background~\cite{PhysRevE.84.061906},   either  imitating  surrounding matrix~\cite{Zhang2013-ju}  or representing  non-mechanical  long-range signaling~\cite{C3SM50838B,Rens2018-yy,Van_Doorn2017-by}. The resultant  competing interactions  generate in the mixed  BS regime stable periodic patterns reminiscent of what is usually observed in  other reinforced fragile  systems~\cite{Xia20001107,Fantilli20091217,Novak2017-nw}.  

We write the dimensionless energy of  our  1D system  in the form   
$
F =  \int_{0} ^1 f dx,
$
where the energy density  has an additive structure  
$
 f(\eps,\eps')=f_1(\eps)+f_2 (\eps').
$  Here   
$\eps(x)=u'(x)$ is  the longitudinal strain,   $u(x)$  is the displacement of point $x$ and prime denotes the derivative.  The term $f_1$ is a single well potential  describing a breakable springs;  in computations  we  use a particular expression 
$
f_1(\eps) = (\eps_0-\eps)^{-2}- 2(\eps_0-\eps)^{-1}, 
$
where $\eps_0-1$ is the reference strain (see Fig. \ref{fig2}), however,  our general results are independent of this choice. With  such    potential, the system's rigidity is sound for sufficiently large stretching  while compression makes the response progressively softer. The second term   in the expression for  $f(\eps,\eps')$  describes  the  bending energy  of the pantographic structure  and, following \cite{alibert:hal-00993920},  we assume that   $
f_2 (\eps')= (\lambda_b^2/2) \eps'^2,
$
where  $\lambda_b$ is  an  \emph{internal} length scale. The resulting model  has the classical GL structure.

We further assume that  the system is loaded in the ``hard'' loading device  which means that  the control parameter is the applied strain $\bar{\eps}$,  so that, for instance,    $ u(0) = - \bar{\eps}/2 ,  u(1) =  \bar{\eps}/2 $.     We also suppose that the  boundaries   of our 'bar' are 'moment free' in the sense  that 
 $u''(0) =  u''(1) = 0.$   Under these conditions  we need to minimize the elastic energy functional $F$. The affine configuration 
$
\bar u (x) = (\bar{\eps}/2)(2x - 1)
$
is always an equilibrium state, however, it is not always stable. To  find the  instability threshold we   study a linearized problem  involving the displacement perturbation $s(x)=  u (x) -\bar u (x)$. The problem reduces  to finding  nontrivial solutions of the linear equation  
\begin{equation}
\label{eq2}
 -\lambda_b^2 s'''' +  \de^2 f_1 (\bar{\eps}) s'' = 0,
\end{equation}
where  $ \de^2 f_1 (\bar{\eps})= \partial^2 f_1/\partial \eps^2\vert_{\bar \eps}$ and the boundary conditions are: $s(0) = s(1) = s''(0) = s''(1) = 0.$ The   non-affine modes  $\sim \sin(n\pi x)$  appear    at $\bar{\eps}$ solving  the characteristic equation
\begin{equation}
\label{eq22}
	\de^2 f_1(\bar{\eps})   = - \lambda_b^2 (n\pi)^2. 
\end{equation}

The analysis of \eqref{eq22} for our choice of the function $f_1$  shows that  the  affine configuration  is locally stable for sufficiently small $\bar{\eps} \leq  \bar{\eps}_c^*$ and for sufficiently large $ \bar{\eps} \geq  \bar{\eps}_c^{**}$. Both direct (affine-non-affine) and return (non-affine-affine)    instabilities are of long wave nature  with  the same critical wavelength $n_c = 1.$  Note that  $ \bar{\eps}_c^{*} \leq  \bar{\eps}_c^{**}\leq  \bar{\eps}_c$,  where  $  \bar{\eps}_c$ is the  spinodal limit satisfying $\de^2 f_1(  \bar{\eps}_c) = 0$,   so the  non-affine configurations  are located inside the  \emph{concavity}  domain of the  potential $f_1$. 

We associate  stable affine states at $ \bar{\eps} \geq \bar{\eps}_c^{**}$ with S-dominated regimes and  at $\bar{\eps} \leq  \bar{\eps}_c^*$ -- with B-dominated regimes. In S regimes the affine character  of the deformation is secured by the presence of  classical elasticity. The latter  becomes destabilizing  in the B regimes where the affinity is  safeguarded by the  (non-classical) gradient elasticity. 
 
\begin{figure}[h]
\includegraphics[scale=0.46]{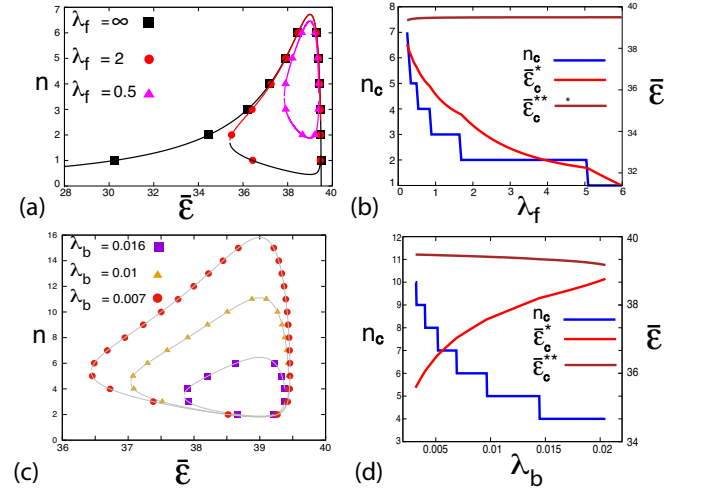}
\caption{Critical strains  $(\bar{\eps}_c^*, \bar{\eps}_c^{**})$ and wave numbers $n_c$ for the  stretched bar on an elastic foundation: (a-b) varying parameter  $\lambda_f$  at fixed  $\lambda_b=0.0167$,
(c-d) varying parameter $\lambda_b$ at  fixed $\lambda_f=0.35$. }
\label{fig3}
\end{figure}
 
To  account for nonlocal  interactions in the realistic biological systems we now  embed our  floppy frame into an elastic environment. To this end we introduce linear coupling of the  GL system with a   pre-stretched background \cite{Truskinovsky1996-xy, Ren:2000aa}. The anti-ferromagnetic effect of the elastic environment will then compete with the ferromagnetic effect of the bending term in the energy  and the resulting microstructures can be more complex. 

We need to consider the dimensionless energy   \cite{Vainchtein1999407}
\begin{equation}
\label{eq222}
F =  \int_{0} ^1 [ f (\eps,\eps') +(1/(2\lambda_f^2))(u - \bar u(x))^2 ]dx, 
\end{equation} 
where  $\lambda_f$ is the \emph{external} length scale characterizing the relative size of the embedding matrix.  Since the  order parameter is $u'(x)$,  the account of   environmental elasticity  brings  implicit nonlocality into the conventional structure of a GL theory~\cite{Ren:2000aa}.

The linear instability condition for the affine state in the model \eqref{eq222} takes the form 
\begin{equation}
\label{eq2222}
 \lambda_b^2 (n\pi)^4 + \de^2 f_1(\bar{\eps})(n\pi)^2 +1/ \lambda_f^2 = 0.
\end{equation} 
One can show that the   redressed upper  $  \bar{\eps}_c^{**}$ and lower $ \bar{\eps}_c^* $  critical strains    correspond again to the same  critical wavelength, however,    $n_c$ can now take arbitrary large values.  In   Fig.  \ref{fig3}  we illustrate the resulting dependence of the critical parameters on  dimensionless lengths $\lambda_b$ and $\lambda_f $. 

Note that the re-entry structure of the bifurcation, found in the local GL system,  persists at  moderate  nonlocality, however,  the non-affinity domain disappears when the internal ($\lambda_b$) and external ($\lambda_f $)  length scales become comparable. The parametric dependence of the critical wavelength takes the form of a staircase which suggests that particular patterns  are robust. 

\begin{figure}[ht]
\vspace{3mm}
\includegraphics[scale=.37]{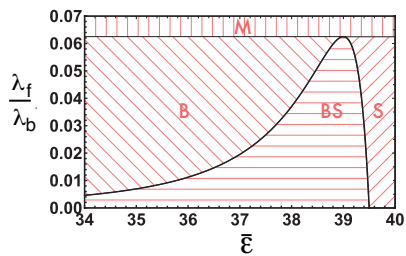}
\caption{ Schematic regime diagram:    (M)  matrix dominated phase;  (B) bending dominated phase ;   (S) stretching  dominated phase, (BS)  mixture phase. }
\label{fig4}
\end{figure}

The obtained results can be summarized in the form of a  regime  diagram. For infinite system, where one can disregard the discreteness of the problem, we can write an approximate equation for  the  critical strain in the form 
\begin{equation}
\label{eq111}
\de^2 f_1(\bar{\eps}_c)=  2  (\lambda_b/\lambda_f ).
\end{equation}
The  solution of \eqref{eq111}  can be used as a rough description of the  boundary delineating the pure  B  and  S  regimes  from the mixed BS regime. In the ensuing diagram (see   Fig. \ref{fig4}) the applied strain $\bar{\eps}$ plays the role analogous to the  cross-linker density, while the ratio  $\lambda_b/\lambda_f$ characterizes the stabilizing  strength of the elastic environment. In the matrix-dominated (super-critical) regime M the deformation is always affine.

\begin{figure}[h!]
 \includegraphics[scale=0.52]{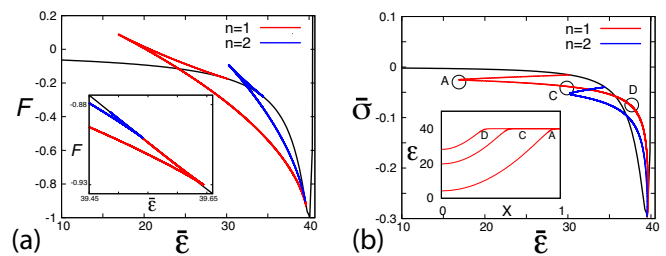}
 \caption{Two lowest energy branches in the problem without elastic foundation: (a)  elastic  energy, (b)  overall stress-strain relation. Black line --  trivial homogeneous branch. Inset in (a) is a zoom  on  the domain where the non-affine  BS  branches merge with the affine S branch. Inset in (b) shows the strain profiles in points A, C and D. Here  $\lambda_b=0.0167$.}
\label{fig5}
\end{figure}

To explore the   structure of the  nonlinear energy minimizing configurations  we need  to solve the   equation 
\begin{equation}
\label{eq11}
-\lambda_b^2 u'''' + \de^2 f_1 (u') u'' - (1/\lambda_f^2) (u - \bar u(x)) = 0.
\end{equation}
 In the limiting case  $\lambda_f=\infty$  the nonlocality is absent and  the  energy  minimizers have the basic  GL structure with  a  single domain boundary  separating the phase where the springs are broken, and the   elasticity is of B type,    from the  phase   where they are intact, and the dominating elasticity is of the S type.  Other equilibrium branches, describing more complex mixtures  of such  phases, have   higher energy, see Fig. \ref{fig5} where we show the equilibrium energy $F(\bar{\eps})$ and the macro stress $\bar{\sigma}(\bar{\eps})= dF(\bar{\eps})/d\bar{\eps}$. 
 
The ensuing two phase configurations, however,  are far from being conventional. Consider, for instance,  a stretching  loading protocol  originating in the homogeneous B phase, and assume that the system always remains in the ground state.  The nucleation of the   S phase takes place discontinuously with the formation of  the configuration   A ( see Fig. \ref{fig5}b)   exhibiting a localized   front separating the \emph{affine}  S phase   and the \emph{non-affine} B phase.  As the applied strain  $\bar{\eps}$ increases, the  homogeneous S phase proliferates with  successive springs reconnecting. During this process the shrinking B phase maintains a particular pattern of non-affinity (see  configurations C and   D in Fig. \ref{fig5}b). The S phase  finally takes over through a discontinuous event of the final annihilation of the B phase.  From the perspective of nonlinear stability theory we observe here a  typical   'isola' bifurcation of re-entry type \cite{dellwo1982}. 
  
\begin{figure}[h!]
\includegraphics[scale=0.51]{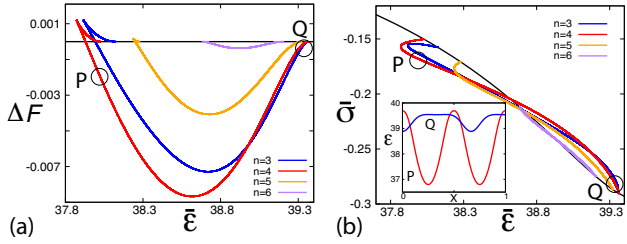}
\caption{Lowest energy branches   in the problem with  elastic environment  \eqref {eq222}:   (a) energy difference between the actual configuration   and the  homogenous configuration; (b) the associated  stress-strain relations. The inset shows  strain profiles for the branch with $n=4$ and $n=3$ corresponding to the  points P and Q. Parameters:  $\lambda_b=0.0167$ and $\lambda_f =0.45$. }
\label{fig6}	
\end{figure}

Consider now  the case when the elastic environment is present ($0< \lambda_f <\infty$).  The bending energy term  then favors coarsening while  the nonlocal  term  drives the refinement of the microstructure   and the ensuing competition  leads to the formation of BS mixtures with more complex geometry. In Fig. \ref{fig6}   we show the typical configuration of the  low energy   branches. Note that the topological structure  of the micro-configuration  changes along the global minimum path: we observe a switch from a configuration with four to a configuration with three BS interfaces. 

 As the total strain  $\bar{\eps} $  increases beyond the point P,  the homogeneous B phase loses stability  which leads to collective  nucleation of the periodically placed islands of the affine S phase  while the remaining B phase becomes non-affine. With further increase of $\bar{\eps}$,  the  islands of  S phase  grow in size, see point Q, and eventually B phase completely disappears. 

We now return to the observation that  in our simple tests the deformation in the   \emph{pure}  S and B phases was affine;  the non-affine response was observed  only in  the BS (mixed) phase.   We recall  that   in experiments  involving fibrous disordered networks,  the non-affinity of the deformation was found in the whole range of the B-dominated elastic response~\cite{RevModPhys.86.995}.   These observations can be explained by  the  fragility  of \emph{affine response} in  B phase while it is robust in  S phase. 
\begin{figure}[ht]
\includegraphics[scale=.31]{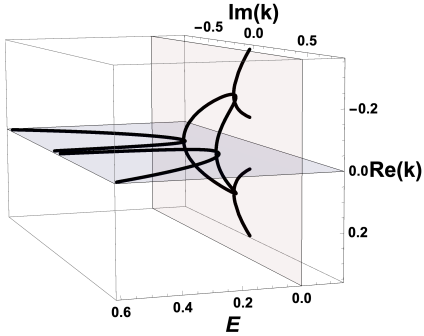}
\caption{Complex roots of the characteristic equation  when  $\lambda_b/\lambda_f=0.03$. The plane $Re$ $k =0$  is highlighted by blue.}
\label{fig8}
\end{figure}

Indeed, consider again the  linear  modes $ \sim \exp{(ikx)}$ superimposed on a homogeneous solution of \eqref{eq11}. The normalized wave numbers $k$ must satisfy the characteristic equation 
$
k^4+E k^2+(\lambda_b/\lambda_f)^2 =0,
$
where for convenience we now introduced directly  the tangential elastic modulus 
$E( \bar \eps)=\partial^2f_1( \bar \eps)$. The complex solutions of this equation are  shown in Fig.  \ref{fig8}.

Note that in  S phase the roots $k$ are  purely imaginary. They describe exponential decay of the local mechanical perturbations and are  characteristic for systems with affine response. Instead, in  B phase  the characteristic wave numbers are real and the  perturbations spread over the whole system signaling non-affine response. In the crossover  range, where $E<0$ and  the scales  $(-E)^{1/2}\lambda_f$ and $(\lambda_b\lambda_f)^{1/2}$ are  comparable (see  \eqref{eq111} for the more precise characterization), the wave numbers are complex and the response is mixed.

\begin{figure}[ht]
\includegraphics[scale=.24]{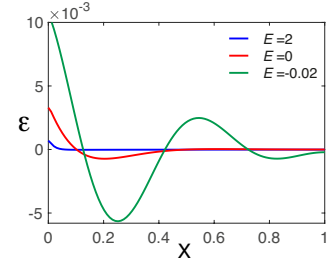}
\caption{Strain profiles appearing in response  to a localized  force distribution  described by the function  $g(x) = 0.1\sech(150(x-0.001))^2$. Parameters: $\bar\epsilon=0$, $\lambda_b=0.0167$ and $\lambda_f=0.5$.}
\label{fig7}
\end{figure}

To further support these observations  we again  linearize   \eqref{eq11}  but now  impose a localized perturbations $g(x)$: 
\begin{equation}
\label{eq12}
-\lambda_b^2 u'''' + Eu'' - (1/\lambda_f^2) (u -\bar u(x) -g(x)) = 0. 
\end{equation} 
The response to such force distribution, when it is   applied near one of the ends of the bar,    is illustrated  in Fig. \ref{fig7}. We see an almost unperturbed affine response in the S phase, a limited penetration of the perturbation in the BS phase and a  markedly non-affine  global response in the B phase.

To show that these observations  are not conditioned to the case with elastic environment, consider  a linearized problem for a bar with $\lambda_f =\infty$ which  is clamped on one side and loaded on the other side by a force $h$.  Suppose that the bending rigidity $\lambda_b$ is sufficiently small, so that in  S phase we can neglect bending and relax the clamping boundary condition. Under these assumptions  the problem reduces to   solving the  equation $u''=0$ with the boundary conditions $u(0)=0, E u'(1)=h$.  The resulting response is   affine:  $\eps=h/E$. Now consider  B phase, where  the stiffness $E$ can be neglected  and the equilibrium equation is  $u'''' =0$,  while  the boundary conditions are   $u(0)=0,u'(0)=0,u''(1)=0,  -\lambda_b^2 u'''(1)=h$. The solution of this boundary value problem is globally inhomogeneous:  $\eps=(h/\lambda_b^2)x(1-x/2)$. 

To conclude, we presented a prototypical \emph{continuum} model of the BS transition. The proposed theory  describes  the peculiar nucleation  and propagation   of S-dominated  domains inside a bar with B-dominated elasticity.   It also explains  the fundamental non-affinity of the B-phase and rationalizes the observed heterogeneity of the mixed BS phase.   To make the model more biologically relevant it will be necessary to   account  for the  fact that  in cellular systems the BS transition  can be also driven  actively  ~\cite{Bertrand23102012,Alvarado:2013aa}.   

We thank P. Recho and P. Ciarletta for helpful discussions. The work was supported by the grants  ANR-18-CE42-0017 (OUS) and ANR-10-IDEX-0001-02 PSL (LT). 
%


\end{document}